\documentclass[prb,preprintnumbers,amsmath,amssymb]{revtex4}
\usepackage{graphicx}
\usepackage{epsfig}
\usepackage{amsmath,amscd}
\usepackage{psfrag}

\begin{document}

\title{Challenges and opportunities for applications of unconventional superconductors.}
\author{Alex Gurevich.}
\affiliation{Department of Physics, Old Dominion University, Norfolk, Virginia 23529; e-mail: gurevich@odu.edu.}

\begin{abstract}

Since the discovery of high-$T_c$ cuprates the quest for new superconductors has shifted toward more anisotropic, strongly correlated materials with lower carrier densities and competing magnetic and charge density wave orders.  While these materials features enhance superconducting correlations, they also result in serious problems for applications at liquid nitrogen (and higher) temperatures and strong magnetic fields, so that such conventional characteristics as the critical temperature $T_c$ and the upper critical field $H_{c2}$ are no longer the main parameters of merit. This happens because of strong fluctuations of the order parameter, thermally-activated hopping of pinned vortices and electromagnetic granularity, as has been established after extensive investigations of cuprates and Fe-based superconductors. In this paper I give an overview of these mechanisms crucial for power and magnet applications and discuss the materials restrictions which have to be satisfied in order to make superconductors useful at high temperatures and magnetic fields. These restrictions become more and more essential at higher temperatures and magnetic fields, particularly for the yet-to-be-discovered superconductors operating at room temperatures. In this case the performance of superconductors will be limited by destructive fluctuations of the order parameter so higher superfluid density and weaker electronic anisotropy which reduce these fluctuations can become far more important than higher $T_c$.

\end{abstract}

\maketitle

\section{Introduction.}

Making predictions in superconductivity, particularly on new materials or the materials requirements for applications of existing or putative room-temperature superconductors (RTS),  has never been rewarding to the point that "doing the opposite" has often worked better than following the conventional wisdom and established models \cite{matthias,berlincourt}.  Yet the important lessons of unprecedented research and development of unconventional superconductors in the last 20-30 years have changed the perception of what is important for applications at high temperatures and magnetic fields. In this paper I discuss some of these lessons which may be helpful for the ongoing quest for new superconductors.

Incidentally, the trend of optimistic predictions was started by the discoverer of superconductivity, Kamerlingh Onnes, who was the first to recognize the advantages of superconducting magnets as the only enabling technology capable of generating the dc magnetic field of 10 tesla which could not be achieved by resistive magnets \cite{onnes}.  However, this idea had to be put aside for a while because all type-I superconductors known before 1930-1940 went to the normal state at  low magnetic fields $H<0.2$ tesla limited by the thermodynamic critical fields $H_c$ of these materials \cite{berlincourt}. The fulfillment of Onnes's vision took nearly 50 years and many scientific and technological breakthroughs, including the discoveries of the Meissner effect \cite{meissner} and type-II superconductors \cite{shubn}, and the development of the London \cite{london}, Ginzburg-Landau (GL) \cite{gl} and Bardeen-Cooper-Schrieffer (BCS) \cite{bcs} theories which showed that superconductivity is a phase-coherent condensate of Cooper pairs glued by phonons. These advances eventually lead to the Abrikosov theory of a lattice of quantized vortices \cite{aaa} which explained how type-II superconductors can withstand high magnetic fields up to $H_{c2}\gg H_c$. It was then realized that the upper critical field $H_{c2}$ above which the type-II superconductivity disappears can be significantly increased by alloying the material with nonmagnetic impurities \cite{gork,whh}.

By 1986 many type-II superconductors, such as Nb-Zr, Nb-Ti, A-15 compounds (Nb$_3$Sn, V$_3$Si, etc.), Chevrel phases (PbMo$_6$S$_8$), with $H_{c2}\simeq 10-60$ tesla and $T_c\simeq 9-23$ K had been discovered, and Nb-Ti and Nb$_3$Sn became the materials of choice for superconducting magnets and medical MRI machines \cite{matthias,berlincourt,natap,dcl}. All these materials are conventional superconductors with the s-wave symmetry of the Cooper pairs described by the Eliashberg theory which generalized the BCS model by taking into account strong electron-phonon interaction \cite{eli1,eli2,eli3,eli4}.  At that time the superconducting critical temperature $T_c$ and the upper critical field $H_{c2}$ were regarded as the main parameters of merit for magnet applications at the liquid helium temperature, 4.2 K. Making superconductors useful involved incorporation of structural defects in a material in order to pin vortices and prevent their dissipative motion under the action of flowing current, which otherwise would cause electric resistance below $H_{c2}$ and $T_c$. Because stronger pinning allows a superconductor to carry larger non-dissipative critical current density $J_c(T,H)$ at high magnetic fields, materials optimization involved incorporating as many extended materials defects and impurities as possible to maximize $J_c$ and $H_{c2}$ without significant degradation of $T_c$ \cite{ce}. Finally, composite wires were produced by embedding thin superconducting filaments into a flexible metallic matrix to provide thermal quench stability, low ac losses and good mechanical properties \cite{wilson}. This approach works for most conventional superconductors such as Nb-Ti ($T_c = 9$ K) and Nb$_3$Sn ($T_c = 18 $ K). More recently, alloying the two-band BCS superconductor MgB$_2$ has increased $H_{c2}(0)$ from $3-5$ tesla to 40-70 tesla \cite{mgb2a,mgb2b,mgb2c}, resulting in the development of magnet conductors \cite{mgb2cond}.

The success of this approach is based on two fundamental features of conventional superconductors in which the symmetry of the order parameter is not lower than the symmetry of the Fermi surface. First, the coherence length $\xi_0 \simeq\hbar v_F/2\pi k_B T_c$ which quantifies the size of Cooper pairs is much greater than the mean electron spacing $r_s$. In a good metal with a large Fermi velocity $v_F$, it is the strong overlap of Cooper pairs which provides the superconducting phase coherence and weak sensitivity of $T_c$ to nonmagnetic impurities and extended crystalline defects such as dislocations and grain boundaries. Another important feature is that fluctuations of the order parameter are negligible because of the smallness of the Ginzburg parameter $Gi = 0.5(k_BT_c/H_c^2\xi^2\xi_c)^2$ proportional to the squared ratio of the thermal energy $k_BT_c$ to the condensation energy $H_c^2\xi^2\xi_c/8\pi$ in the volume occupied by the Cooper pair \cite{fluct}. Here $\xi$ is the in-plane coherence length, $\xi_c=\xi/\gamma$ is the coherence length along the c-axis, and the anisotropy parameter $\gamma=(m_c/m)^{1/2}$ is defined by the ratio of effective masses along the c-axis ($m_c$) and in the ab plane ($m$) in a uniaxial superconductor \cite{blatter,ehb}. In conventional superconductors $Gi$ varies from $\sim 10^{-11}$ for Nb ($T_c=9.2$ K) to $\sim 10^{-4}$ for a two-band MgB$_2$ with $T_c=40$ K. The materials parameters which control $Gi$ become more transparent if $H_c^2=\phi_0^2/8\pi^2\xi^2\lambda^2$ is expressed in terms of the magnetic London penetration depth, $\lambda = (mc^2/4\pi n_se^2)^{1/2}$ in the clean, single band limit, where $n_s$ is the superfluid density equal to the carrier density at $T=0$, and $\phi_0=\pi\hbar c/e$ is the flux quantum \cite{book}:
\begin{equation}
Gi=2\left(\frac{m\gamma k_BT_c}{\pi\hbar^2n_s\xi_0}\right)^2.
\label{Gi}
\end{equation}
In moderately anisotropic superconductors $Gi\propto m^2T_c^4\gamma^2n_s^{-8/3}$ increases strongly as $T_c$, $\gamma$ and $m$ increase or $n_s$ decreases. For very anisotropic layered materials, $Gi=2mk_BT_c/\pi d\hbar^2n_s$ where $d$ is the spacing between layers \cite{blatter}. If fluctuations are negligible, $J_c(H)$ is controlled by pinning of vortices and usually follows the phenomenological field dependence $J_c(H) \propto b^{-\alpha}(1-b)^\beta$ where  $\alpha=0.3-1$, $\beta = 1-2$, and $b=H/H_{c2}$ \cite{ce}. Since $J_c(H)$ vanishes at $H_{c2}$, adding nonmagnetic impurities shortens the mean free path $\ell$ in the dirty limit $\ell < \xi_0$ and extends the field region $H<H_{c2}\simeq\phi_0/2\pi\xi_0\ell\propto T_c$ \cite{gork,whh} where superconductors can carry weakly dissipative currents. For conventional superconductors, both $H_{c2}$ and $J_c$ scale with $T_c$ \cite{ce}, so the higher $T_c$ the better. In turn, the search for materials with higher $T_c$ was guided by the famous "Matthias rules": 1. Transition metals are better than simple metals, 2. Peaks in the electron density of states at the Fermi surface are good, 3. High crystal symmetry (especially cubic) is good, 4. Stay away from oxygen, magnetic and dielectric states \cite{matthias}.

The discoveries of the Chevrel phases \cite{chevrel}, heavy fermions \cite{hf1,hf2}, and organic superconductors \cite{orgsc1,orgsc2} gave first indications that the approach outlined above may not work for superconductors with small (nanoscale) $\xi_0$, non s-wave pairing, and strong vortex fluctuations in quasi-one dimensional or layered materials. These features, first regarded as exotic and not relevant to practical conductors, were eventually recognized as being among the key issues for applications at 77 K, triggered by the groundbreaking discovery of high-$T_c$ cuprates \cite{bedmul}. The initial enthusiasm about powerful high-field magnets, motors, generators and transmission lines working at liquid nitrogen temperatures ($77$ K) was based on a belief that the high $T_c$ of YBa$_2$Cu$_3$O$_{7-x}$ ($T_c = 92$ K) and (Bi,Pb)$_2$Sr$_2$Ca$_2$Cu$_3$O$_x$ ($T_c = 110$ K) would somehow assure high-field conductors. However, applications at 77 K have turned out to be much more challenging than at 4.2 K, regardless of the values of $T_c$ and $H_{c2}$. Moreover, the fascinating physics and materials science behind high $T_c$ and $H_{c2}$ in  cuprates and the recently discovered Fe-based superconductors (FBS) \cite{fbs,fbs1,fbs2,fbs3,fbs4,fbs5,fbs6} can also result in obstacles for applications. Making such materials useful involves compromises among conflicting requirements, defining the parameters of merit depending on the operating conditions and also on the specific application.

I will discuss superconductors for power and magnet applications at high temperatures and magnetic fields for which $T_c$ has been historically the most captivating parameter of merits, implying that "better" superconductors have higher $T_c$. The discovery of cuprates for the first time enabled applications at 77 K but it also revealed that: 1. Because of strong thermal fluctuations of vortices, $J_c(H)$ vanishes at the irreversibility field $H^*(T)$ which can be much smaller than $H_{c2}(T)$. 2. Current-blocking grain boundaries in cuprates \cite{gbhts} and FBS \cite{gbfbs} significantly reduce $J_c$ in polycrystals.  The bad combination of these two issues is one of the most serious materials challenge for power and magnet applications at 77 K. These problems would be even more severe for RTS operating at room temperatures for which the current-carrying capability will be mostly limited by fluctuations of the order parameter and thermally-activated hopping and reconnection of soft vortex segments rather than by the high pairing temperature at which the Copper pairs form. In this case high superfluid density and low electron mass anisotropy become the key normal state materials properties which reduce thermal fluctuations and enable phase coherence and superconducting currents.

\section{Problems with strongly correlated layered materials}

The field dependence of $J_c(T,H)$ is of major importance for magnet applications. The value $J_c(4.2 K, 5 T)\simeq 0.5$ MA cm$^{-2}$ is characteristic of Nb47wt$\%$Ti alloys with $T_c = 9$ K and $H_{c2}(4.2 K) = 12$ tesla used in magnets \cite{wilson}.  Many superconductors have $H_{c2}$ much higher than $H_{c2}(0)=15$ tesla of NbTi because they have shorter $\xi_0$ and can sustain stronger fields up to $H_{c2}(0)\simeq \phi_0/2\pi\xi_0^2$ at which the spacing between vortices $(H/\phi_0)^{1/2}$ becomes of the order of the diameter of nonsuperconducting vortex cores $\simeq 2\xi_0$.  Very high $H_{c2}$ of the cuprates and FBS result from their small $\xi_0 = \hbar v_F/2\pi k_BT_c$, either due to high $T_c = 90-130$ K in cuprates or smaller $v_F$ in semi-metallic FBS. The values of $H_{c2}(0) > 100$ tesla extrapolated from low-field measurements near $T_c$ often exceed the BCS paramagnetic limit $H_p$ at which the Zeeman energy equals the binding energy of Cooper pairs with antiparallel spins, $H_p$[tesla] $= 1.84T_c$[K] \cite{ashc2,tarantHc2}.  \textbf{Figure 1} shows $H_{c2}(T)$ and $H^*(T)$ for some low-$T_c$ superconductors, cuprates and FBS.

\begin{figure}
\epsfxsize28pc
\centerline{\epsfbox{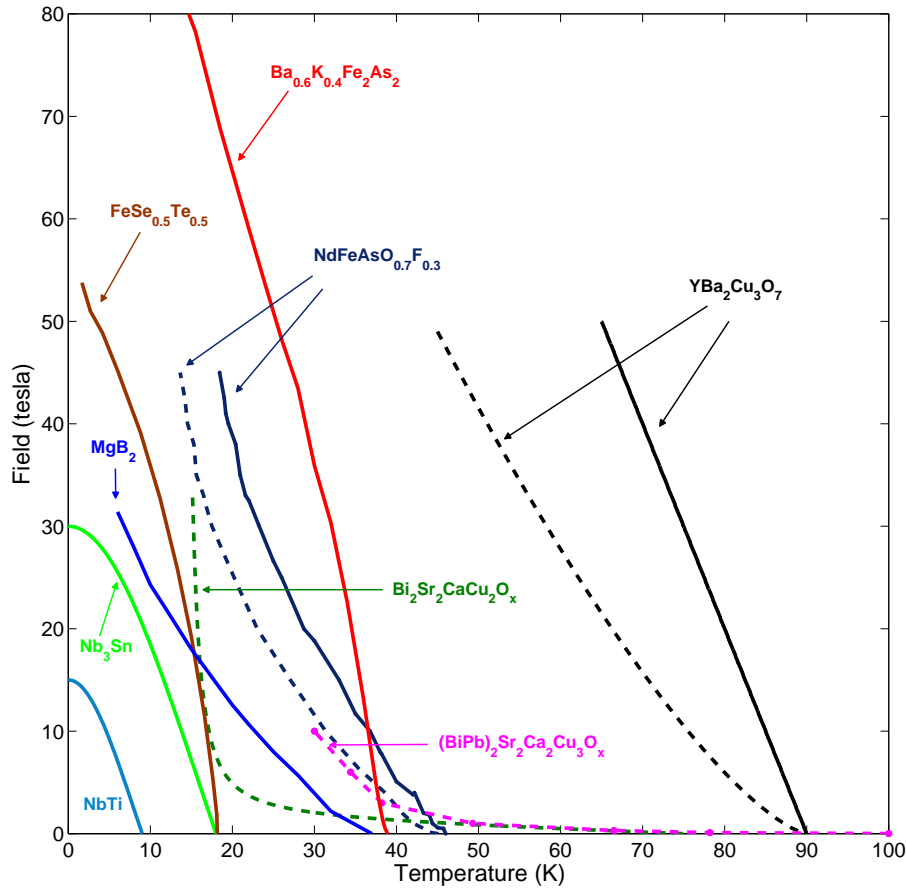}}
\caption{Comparative $H-T$ phase diagram for representative cuprates, FBS, and conventional superconductors,  where the solid and dashed lines show $H_{c2}(T)$ and $H^*(T)$ parallel to the c-axis, respectively. In the range of fields 5 tesla $< H <$ 70 tesla important for magnets, the $H_{c2}(T)$ curves for most FBS with $T_c > 20$ K are clustered between $H^*(T) $ of the layered Bi$_2$Sr$_2$CaCu$_2$O$_x$ and $H^*(T)$ for the least anisotropic YBa$_2$Cu$_3$O$_{7-x}$. Here $H^*(T)$ for the layered Bi$_2$Sr$_2$CaCu$_2$O$_x$ ($T_c\approx 75$ K) and Bi$_2$Sr$_2$Ca$_2$Cu$_3$O$_x$ ($T_c\approx 108$ K) are much smaller than their respective $H_{c2}(T)$ which have slopes $dH_{c2}/dT\simeq 2$ tesla K$^{-1}$ at $T_c$ (not shown here). The difference between $H^*$ and $H_{c2}$ for NdFeAsO$_{0.7}$F$_{0.3}$ at 20-30 K is smaller than the difference between $H^*$ and $H_{c2}$ for YBa$_2$Cu$_3$O$_{7-x}$ at 77 K, which reflects the diminishing role of vortex fluctuations at lower $T$. The less anisotropic Ba$_{0.6}$K$_{0.4}$Fe$_2$As$_2$ with $1 < \gamma (T) < 2$ and $T_c = 38$ K has a higher $dH_{c2}/dT$ than NdFeAsO$_{0.7}$F$_{0.3}$ with $\gamma(T) \approx 4-8$ and $T_c = 42$ K \cite{tarantHc2}, so the Ba$_{0.6}$K$_{0.4}$Fe$_2$As$_2$ polycrystalline conductors which also exhibit weaker GB problem \cite{ast10} could be superior at 20 K. The data are reproduced from Ref. \cite{agnat} with the addition of recent $H_{c2}$ data for FeSe$_{0.5}$Te$_{0.5}$ and Ba$_{0.6}$K$_{0.4}$Fe$_2$As$_2$ from Ref. \cite{tarantHc2}.}
\label{Figure 1}
\end{figure}

In cuprates and to a lesser extent in FSB, $H^*$ is controlled by thermal fluctuations of vortices which weaken pinning and cause giant magnetic flux creep \cite{creep} and electrical resistance well below $H_{c2}$. In this case a superconductor can carry weakly dissipative current only below the irreversibility field $H^*(T)$ at which $J_c(H)$ vanishes. At higher fields $H^* < H < H_{c2}$ cuprates and FBS behave as poor metals and their high $H_{c2}$ become irrelevant for applications. The dissipative domain  $H^* < H < H_{c2}$ widens significantly in anisotropic materials. For instance, \textbf{Figure 1} shows that, for layered cuprates, $H^*(T)$ (dashed line) can be much lower than $H_{c2}(T)$ (solid line).  The main reason is the loss of shear rigidity of vortex structures above the field $H_m(T) \simeq (5-8)\cdot 10^{-3}H_{c2}(0)(T_{c}/T  -  1)^2/Gi \simeq H^*$ \cite{blatter,ehb} which makes pinning ineffective. The field $H_m$ at which the hexagonal vortex lattice melts, decreases dramatically in anisotropic materials with high $T_c$, low $n_s$ or heavy electron mass $m$.

There is no general relation between the melting fields $H_m(T)$ of an ideal vortex lattice in a defect-free superconductor and the irreversibility field $H^*(T)$ at which thermal depinning of a vortex glass structure disordered by materials defects occurs \cite{blatter,ehb}. Because pinning hinders thermal wandering of vortices, one can only conclude that $H_m<H^*<H_{c2}$ and $H_m$ is the lower bound of $H^*$ for weak sparse pins. The theory of $H^*$ is not only much more complicated than $H_m$ but is also depends on many uncertain materials details and particular mechanisms of interaction of vortices with defects, shape and spatial distribution of pinning centers etc. For this reason I will be using here the well established and simpler theory of $H_m$ to discuss the materials parameters which limit the high-field performance of superconductors. In YBa$_2$Cu$_3$O$_{7-x}$ even extremely strong pinning only weakly affects $H^*$ which turns out to be close to $H_m(T)$.

In conventional superconductors $H_m \approx H_{c2}$, but even the moderately anisotropic YBa$_2$Cu$_3$O$_{7-x}$ with $Gi \sim 10^{-2}$ has $H^*(77 K)\simeq 7-10$ tesla $\simeq 0.5H_{c2}(77 K)$, while the extremely anisotropic layered (Bi,Pb)$_2$Sr$_2$Ca$_2$Cu$_3$O$_x$ with $Gi \sim 1$  has much lower $H_m(77K)\simeq H^*(77 K) \ll H_{c2}(77 K)$.  FSB have $Gi\sim 10^{-5} - 10^{-2} $ \cite{fbs4}, so NdFeAsO$_{1-x}$F$_x$ and Ba$_{0.6}$K$_{0.4}$Fe$_2$As$_2$ have $H^*(T) > 30$ tesla at 20-30 K where FBS outperform both MgB$_2$ and (Bi,Pb)$_2$Sr$_2$Ca$_2$Cu$_3$O$_x$, as shown in \textbf{Figure 1}.

It appears that low $H^*$ at $77$ K in optimally doped cuprates cannot be significantly increased by refinements of chemical phase composition and incorporation of appropriate defects to pin strongly fluctuating vortices. In fact, this program has been implemented for YBa$_2$Cu$_3$O$_{7-x}$ in which dense arrays of oxide nanoparticles were incorporated to pin every $5-10$ nm of vortex lines \cite{y1,y2,y3,y4,y5,y6}.  As a result, $J_c(77 K)$ of $1-2\ \mu$m thick YBa$_2$Cu$_3$O$_{7-x}$ films in the second generation conductors was pushed up to $\simeq 20-30\%$ of the depairing current density $J_d=c\phi_0/12\sqrt{3}\pi^2\lambda^2\xi$ - the maximum current density a superconductor can carry in the Meissner state.  However, it turned out that pinning nanostructures, which enhance $J_c$ so effectively at low fields do not really increase $H^*$ by preventing thermal wandering of vortices. Even for YBa$_2$Cu$_3$O$_{7-x}$ films with highest values of $J_c(77 K)\simeq 7-8$ MA cm$^{-2}$ in self field, the maximum $H^*(77 K)\simeq 10-11$ tesla \cite{y1,y2,y3,y4,y5,y6} is not much higher than the melting field $H_m(77 K)\approx 9$ tesla observed by calorimetric measurements on single crystals \cite{Hm}. Both $H_m(77 K)$ and $H^*(77 K)$ are well below $H_{c2}(77 K)\simeq 30$ tesla (see {\bf Figure 1}), indicating that $H^*$ may be limited by intrinsic materials parameters on the scales $\simeq\xi = 2-4$ nm.

Strong pinning of vortices and high $H^*$ are not yet sufficient for applications, which also require long polycrystalline wires. One of the main issues for the cuprates is that grain boundaries (GBs) between misoriented crystallites impede current flow because the GB critical current density $J_g(\theta) = J_0\exp(- \theta/\theta_0)$ drops exponentially as the misorientation angle $\theta$ between neighboring crystallites increases \cite{gbhts}. For cuprates, $\theta_0 \approx  3-5^\circ$ so the spread of misorientation angles $\Delta\theta \simeq 40^o$ in polycrystals can reduce $J_g$ by 2-3 orders of magnitude!  FBS polycrystals also have weak linked GBs \cite{fbs5,gbfbs}, although the observed larger values of $\theta_0\simeq 5-9^\circ$ are definitely beneficial for applications.

\begin{figure}
\epsfxsize15pc
\centerline{\epsfbox{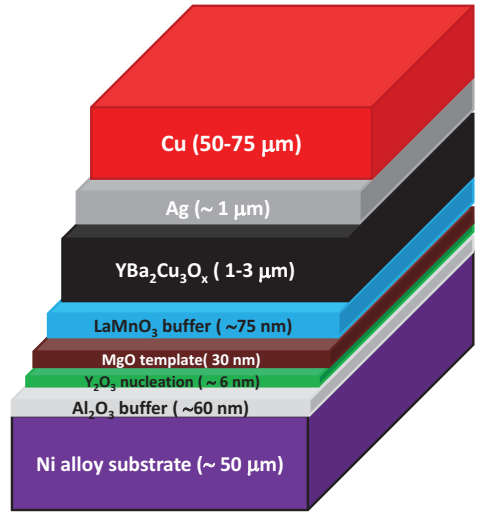}}
\caption{A typical architecture of a coated conductor made by the ion beam assisted deposition \cite{cc}. The YBa$_2$Cu$_3$O$_{7-x}$ film is grown on a textured Ni alloyed substrate with a complex buffer layer structure, which enables replication of the low-angle grain structure of the substrate in YBa$_2$Cu$_3$O$_{7-x}$ and protects it from chemical contamination. The stabilizing layers of Ag and Cu on top of the $1-2 \mu$m thick  YBa$_2$Cu$_3$O$_{7-x}$ film provide thermal quench protection of the tape which are usually few mm wide and 0.1-0.2 mm thick. The current-carrying superconducting film takes 1-2$\%$ of the conductor cross-section, which strongly reduces its averaged current density.  Reproduced from Ref. \cite{agnat}.}
\label{Figure 2}
\end{figure}

Discovered in 1988, the current-limiting GBs in cuprates \cite{gbhts} were immediately recognized as a serious problem for applications because, instead of flowing along the wire, current breaks into disconnected loops circulating in the grains. This problem has been eventually addressed by the coated conductor technology in which the fraction of high-angle GBs with  $\theta > 5-7^\circ$ is reduced by growing YBa$_2$Cu$_3$O$_{7-x}$ films on textured metallic substrates \cite{cc}. \textbf{Figure 2} shows an example of the coated conductor architecture, which has made the idea of YBa$_2$Cu$_3$O$_{7-x}$ "single crystal by the mile" a reality available for power and magnet applications \cite{cc,malozem}. These state-of-the-art coated conductors have several drawbacks: 1. Growing long (hundreds of meters) YBa$_2$Cu$_3$O$_{7-x}$ films and complex buffer layers on textured substrates is much slower and more expensive than the production of conventional multifilamentary wires, 2. Planar coated conductor geometry strongly increases the electromagnetic losses in alternating magnetic fields. 3. The current-carrying YBa$_2$Cu$_3$O$_{7-x}$ film takes only 1-2$\%$ of the conductor cross section, so to make such conductors competitive, $J_c$ of YBa$_2$Cu$_3$O$_{7-x}$ film must be pushed to its limit by the addition of strong pinning centers, e.g., oxide nanoparticles.

The problems outlined above result from generic features of cuprates and FBS which appear common for other unconventional superconductors as well. Both cuprates and FBS are layered materials in which superconductivity primarily occurs on atomic planes (the Cu-O planes in the cuprates and the Fe-As or Fe planes in FBS). FBS comprise 4 main families: ReFeAsO$_{1-x}$F$_x$ with Re = La, Sm, Nd and $T_c$ up to 55 K, (Ba$_x$M$_{1-x}$)Fe$_2$As$_2$ with M = Co, K and $T_c$ up to 38 K,  MFeAs with M = Li, Na and $T_c$ up to 18 K, and  FeSe$_{1-x}$Te$_x$ with $T_c$ up to 18 K \cite{fbs1,fbs2,fbs3,fbs4,fbs5}. Superconductivity in cuprates occurs on doping a Mott antiferromagnetic (AF) insulator \cite{hts1,hts2,hts3,hts4,hts5,hts6,hts7} while FBS become superconducting on doping a parent AF semi-metal \cite{fbs,fbs1,fbs2,fbs3,fbs4,fbs5,fbs6}. The following features of cuprates and FBS cause problems for applications:
\begin{itemize}
\item{High normal state resistivities $\rho_n$, low carrier densities, and low Fermi energies as compared to conventional superconductors. As a result, both FBS and the cuprates have small Cooper pairs with $\xi_0  \simeq 1-2$ nm, but large Thomas-Fermi screening lengths $l_{TF} \simeq \xi_0$.}
\item{Proximity of superconductivity to competing AF states.}
\item{Unconventional symmetry of the Cooper pairs: d-wave in cuprates \cite{hts2} and multiband d-wave or $s_{\pm}$-wave with a sign change of the superconducting gap on disconnected pieces of the Fermi surface in FBS  \cite{hts7,fbst1,fbst2}.}
\item{Large mass anisotropy parameter $\gamma = (m_c/m)^{1/2}$  ranging from $\simeq 1$ to $\simeq 7$ for FBS and from $\simeq 5$ to $>100$ for the cuprates.}
\item{Complex chemical compositions, precipitation of second phases and sensitivity of superconducting properties to local nonstoichiometry.}
\end{itemize}
These materials properties facilitate thermal fluctuations of vortices and electromagnetic granularity, significantly reducing the useful $T-H$ domain in which superconductors can carry currents. As a result, complex and expensive technologies had to be developed to enable applications of YBa$_2$Cu$_3$O$_{7-x}$ at 77 K. {\it These points become even more important as the operating temperature and magnetic field increase, imposing significant restrictions on the normal state electronic parameters of RTS if they would be used at room temperatures.}

\begin{figure}
\epsfxsize15pc
\centerline{\epsfbox{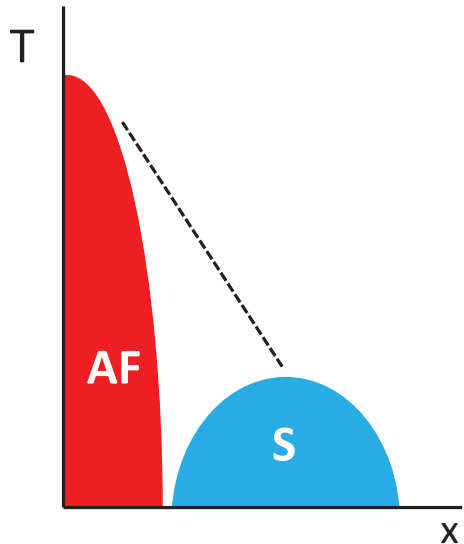}}
\caption{Generic phase diagram of unconventional superconductivity emerging on doping an AF parent state. Here $x$ can be proportional to either hole or electron concentration, or to other external factors such as pressure.}
\label{Figure 3}
\end{figure}

One of the issues for applications results from the generic phase diagram shown in \textbf{Figure 3} where the superconducting dome appears on doping an AF state. This phase diagram is a trademark of unconventional superconductors \cite{dagoto,maple} such as cuprates, FBS, heavy fermions \cite{hf1,hf2} and organic superconductors \cite{orgsc1,orgsc2}. Superconductivity emerging from doped AF states is likely behind the high-$T_c$ values of cuprates and FBS in which the Cooper pairing is mediated by magnetic excitations \cite{hts3,hts4,hts5,hts6,hts7,fbst1,fbst2}, but smaller Fermi energies, high anisotropy, unconventional pairing symmetry, and the proximity to AF states can also cause problems for applications. Indeed, $E_F\simeq 200-300$ meV of optimally doped cuprates and even lower $E_F\simeq 30-100$ meV in FBS are $1-2$ orders of magnitude smaller than $E_F\simeq 2-10$ eV of metallic superconductors. If such common crystal defects as grain boundaries or dislocations locally change the chemical potential by $\sim 0.1 $ eV, it would weakly affect properties of conventional superconductors but cause strong local fluctuations of properties in strongly correlated superconductors with $E_F\sim 0.1$ eV. Moreover, neither FBS-like semi-metals with $E_F=30$ meV = 330 K nor similarly doped semiconductors \cite{sem1,sem2} or fullerides \cite{ful} can be RTS simply because an electron Fermi liquid with $E_F<0.1$ eV turns into a classical plasma at room temperatures.

Given the current trend of searching for new superconductors among anisotropic, strongly correlated materials with low carrier density, non s-wave pairing and competing orders, one can ask the question which was recently posed by Beasley \cite{mac} (see also \cite{agnat}). Suppose that an RTS with $T_c>300$ K has been discovered.  Can it be made useful for applications at 300 K?  More specifically:
\begin{enumerate}
\item{ What are the materials requirements which would enable RTS to work at $T>77$K or even at room temperatures and high magnetic fields?}
\item{To what extent can the performance of RTS be improved by defect nanostructures, assuming that it will not be prohibitively expensive?}
\end{enumerate}
Discussing these issues may look premature given that no RTS has been discovered, while theories of superconductivity in cuprates and FBS do not have the predictive power of the BCS/Eliashberg theory which, in turn, cannot calculate $T_c$ accurately because of the exponential sensitivity of $T_c$ to electron and phonon parameters and insufficient accuracy in the evaluation of the Coulomb pseudopotential \cite{eli2,eli3,eli4}. However, the situation may change for the better at high $T$ and $H$ where the behavior of RTS is mostly controlled by fluctuations and becomes less sensitive to the poorly understood details of non-BCS pairing mechanisms. In this case $T_c$ at which the global phase coherence sets in, and $H^*$ are primarily determined by such normal state properties as $n_s$ and $\gamma$ \cite{mac,agnat}, so the qualitative requirements of useful $J_c(T,H)$ and $H^*(T)$ can at least be formulated.

The importance of phase fluctuations of the order parameter for the reduction of $T_c$ in anisotropic superconductors with low superfluid density was pointed out by Emery and Kivelson \cite{kivels1}. These ideas have been widely used to understand properties of underdoped cuprates \cite{kivels3, tallon}. At the same time, the major role of thermal fluctuations of vortices on $J_c$ and $H^*$ in anisotropic cuprates has also been established \cite{blatter,ehb,tinkh}. Based on these results, one can identify the essential physical parameters, particularly the carrier density and crystal anisotropy, which will limit applications of superconductors at high $T$ and $H$ \cite{mac,agnat}.

The main problem is already apparent from Equation \ref{Gi} which shows that a two-fold increase of $T_c$ yields a 16-fold increase of $Gi\propto m^2T_c^4\gamma^2n_s^{-8/3}$. Thus, high field performance of MgB$_2$ or FBS at 30 K would be affected by fluctuations much weaker than of YBa$_2$Cu$_3$O$_{7-x}$ or more anisotropic cuprates at 77 K. Going to 300 K would increase $Gi$ by $\sim 10^4$ times, in which case the performance of RTS would be mostly limited by fluctuations rather than by Cooper pairing. I will discuss the mechanisms which affect applications at higher $T$ and $H$, assuming the most favorable conditions for pinning and thermal fluctuations. This qualitative analysis underestimates the materials requirements for RTS applications, yet it shows that satisfying even these optimistic conditions is not going to be easy.

\section{What do power applications need?}

The cuprates or FBS can be invaluable for magnets, transmission lines, motors or generators operating at $T>4.2$ K. For example, the second generation coated conductors \cite{cc,cc1,cc2}, melt-textured YBa$_2$Cu$_3$O$_{7-x}$ \cite{meltext}, and Bi-Sr-Ca-Cu-O tapes \cite{tape1,tape2} are already being used in the power grid \cite{malozem}, high-field magnets \cite{htsmagnet}, motors and generators \cite{motors1,motors2}. The younger FBS conductor technology has been recently advancing with impressive rate \cite{fbs5,fbswire}. Magnet applications require conductors with $J_c\simeq 10-100$ kA cm$^{-2}$, preferably weakly dependent on the orientation of \textbf{H} \cite{natap,dcl}.

So far just six superconductors have become commercial magnet materials. These are the conventional superconductors, Nb-Ti, Nb$_3$Sn \cite{natap} and MgB$_2$ \cite{mgb2cond}. Of the cuprates, only three have been used in magnets, ac cables or current leads: Bi$_2$Sr$_2$CaCu$_2$O$_{8-x}$, (Bi,Pb)$_2$Sr$_2$Ca$_2$Cu$_3$O$_{10-x}$ and YBa$_2$Cu$_3$O$_{7-x}$ \cite{dcl,tape1,tape2}. For magnet applications at 77 K, only YBa$_2$Cu$_3$O$_{7-x}$ can be used because layered  Bi-Sr-Ca-Ca-Cu-O cuprates have very low $H^*(77K)<$ 1 tesla, as it is evident from \textbf{Figure 1}. The least anisotropic YBa$_2$Cu$_3$O$_{7-x}$ coated conductors currently provide the only enabling magnet technology at 77 K.

Applications of FBS are limited to temperatures $10-30$ K at which cryocoolers are effective and (Ba$_x$K$_{1-x}$)Fe$_2$As$_2$ or NdFeAsO$_{1-x}$F$_x$ have high $H^*$ and $H_{c2}$ up to 70-80 tesla. Moreover, $H_{c2}$ of (Ba$_x$K$_{1-x}$)Fe$_2$As$_2$ is much less anisotropic than $H_{c2}$ of YBa$_2$Cu$_3$O$_{7-x}$ and NdFeAsO$_{1-x}$F$_x$, which is also useful for magnets. The full extent to which poor grain connectivity in FBS is essential remains unclear, but the first results have shown that the GB problem in FBS is less severe than in cuprates \cite{fbs5,gbfbs,fbswire}. As far as pinning is concerned, FBS films and single crystals can have high and weakly anisotropic  $J_c\sim 1-5$ MA cm$^{-2}$ at 4.2 K  \cite{ast1,ast2,ast3,ast4,ast5,ast6,ast7,ast8,ast9,ast10,ast11}. For instance, Ba$_x$K$_{1-x}$Fe$_2$As$_2$ may be a good magnet material because its high and moderately anisotropic $H_{c2}(20K)\simeq 70$ tesla could make it more useful than the more anisotropic NdFeAsO$_{1-x}$F$_x$ with higher $T_c$, whereas $J_c(5K)$ can reach up to 5 MA cm$^{-2} \simeq 20\% J_d$ and be practically independent of $H$ up to 7 tesla \cite{ast11}. In turn, $H_{c2}$ of FeSe$_{0.5}$Te$_{0.5}$ is twice that of Nb$_3$Sn \cite{a15} at nearly the same $T_c\approx 18$ K, and chalcogenide coated conductors can have $J_c(4.2K)>1$ MA cm$^{-2}$ in self field and $J_c(4,2K,H)>0.1$  MA cm$^{-2}$ in fields up to 30 tesla \cite{ast9}.

It is clear that cuprates and FBS do have many attractive materials properties, but they are only one necessary prerequisite for applications which also include the price, complexity, and environmental footprint of the conductor technology. It is therefore not surprising that so far, cheaper, less anisotropic and more technological materials like Nb-Ti, Nb$_3$Sn, MgB$_2$ or YBa$_2$Cu$_3$O$_{7-x}$ have won the race for magnet applications although many superconductors have much higher $T_c$ and $H_{c2}$. Toxicity of chemical components is also a problem for Hg or Tl-based cuprates with the highest $T_c\approx 132$ K \cite{hts1}. This may also be relevant for the arsenic-containing FBS.

\subsection{Vortex melting field}

Thermal wandering of vortices in anisotropic superconductors makes pinning inefficient at high $T$ and $H$ and causes melting of the vortex lattice at $H=H_m<H_{c2}$. The melting field $H_m$ is controlled by the dispersive line tension of the vortex $\varepsilon_l(k)$ for short-wavelength bending distortions $k\lambda\gg 1$ \cite{blatter,ehb}:
\begin{equation}
\varepsilon_l(k)=\frac{\epsilon_0}{\gamma^2}\ln\frac{1}{k\xi_c}, \qquad
\varepsilon_0=\left(\frac{\phi_0}{4\pi\lambda}\right)^2=\frac{\pi\hbar^2 n_s}{4m}
\label{eps}
\end{equation}
Anisotropy and low superfluid density reduce bending rigidity of a vortex. For instance, $\varepsilon_l\simeq 10^4$ K nm$^{-1}$ in Nb ($\lambda \simeq 40$ nm and $\gamma=1$),  while for YBa$_2$Cu$_3$O$_{7-x}$ with $\lambda(77K)\simeq 400$ nm and $\gamma=5$  \cite{lamh1,lamh2,lamh3,lamh4}, Equation \ref{eps} yields $\varepsilon_l\simeq 5$ K nm$^{-1}$. Most FBS have $\varepsilon_l\simeq (200-460)(1-T^2/T_c^2)$ K nm$^{-1}$ due to larger $\lambda_0\simeq 200-300$ nm but smaller $\gamma$ \cite{lamf1,lamf2,lamf3}.  The Bi-based cuprates in which weakly coupled pancake vortices on Cu-O planes fluctuate strongly, have much lower $\varepsilon_l < 1$ K nm$^{-1}$ at $77$ K \cite{blatter}. Thus, vortices in Nb form the rigid Abrikosov lattice, but in cuprates at $77$ K thermal fluctuations can displace vortex segments of length $L\simeq 10-20$ nm larger than the intervortex spacing at fields $H\sim \phi_0/L^2$, smearing out the random pinning potential and reducing $J_c(T,H)$.

Low bending rigidity of vortices increases the amplitude $u$ of their fluctuations, causing melting of the vortex lattice at $\langle u^2(T,H)\rangle=c_L^2a^2$, where $a=(\phi_0/H)^{1/2}$ is the spacing between vortices and $c_L\approx 0.17-0.15$ is the Lindemann number \cite{blatter,ehb}. The melting field can be estimated by equating the bending energy of the vortex segment of length $L$ to the thermal energy: $\varepsilon_l u^2/L \sim k_BT$. In turn, the energy of tilt distortions is of the order of the energy of shear distortions: $\varepsilon_l a^2/L^2 \sim \varepsilon_0$, so $L\sim a\gamma^2$. Using the GL temperature dependencies, $\varepsilon_0(T)=\varepsilon_0(0)(1-T/T_c) $ yields $H_m = H_0(T_c/T-1)^2$, where $H_0=c_L^4\phi_0\epsilon[\varepsilon_0(0)/k_BT_c]^2$. Calculations based on the nonlocal elasticity theory of vortex lattice yield \cite{blatter,ehb}:
    \begin{equation}
    H_m=H_0\left(\frac{T_c}{T}-1\right)^2, \qquad
    H_0=\frac{\pi^3\phi_0c_L^4}{4\gamma^2}\left(\frac{\hbar^2n_s}{mk_BT_c}\right)^2
    \label{hm}
    \end{equation}
Here $H_m(T)$ has upward curvature near $T_{c}$ where $H_m(T)\propto (1-T/T_c)^2$ is much smaller than $H_{c2}(T)\propto 1-T/T_c$, and crosses over with $H_{c2}(T)$ below $T\simeq T_G\simeq T_c(Gi_c/Gi)^{1/2}$ where $Gi_c=\pi^2c_L^4\simeq 5\cdot 10^{-3}$. The values $Gi\sim 10^{-3}-10^{-2}$ are characteristic of YBa$_2$Cu$_3$O$_{7-x}$, or FBS like Re(O$_{1-x}$F$_x$)FeAs with Re = La, Sm, Nd and chalcogenides. \textbf{Figure 1} shows that in the layered Bi-Sr-Ca-Cu-O crystals, $H^*(T)< 10$ tesla at $T>30-35$ K, well below $T_c$. The reduction of $H_m$  by high anisotropy and low $n_s$ also manifests itself in thermal depinning of vortices from columnar defects considered in the next subsection.

For moderately anisotropic RTS with $Gi\sim 1$, the melting field $H_m$ becomes independent of superconducting parameters in a wide temperature region $T_G\ll T\ll T_c$ where $H_m\simeq \pi\phi_0(\pi c_L^2 n_s/2m\gamma k_BT)^2$ is controlled by the ratio $(n_s/m\gamma)^2$ reduced by smaller superfluid density, crystal anisotropy and bigger effective mass in strongly correlated materials. For $\lambda_0=200$ nm and $\gamma=5$, we obtain $H_m\simeq 1.2$ tesla at 300K even if $T_c\gg 300$ K.  {\it Thus, applications at $10-20$ tesla and 300 K would require either an isotropic RTS with $\gamma=1$ or an RTS with higher $n_s$ to reduce the London penetration depth to the level of $\lambda_0 < 100$ nm  characteristic of such conventional superconductors as Nb$_3$Sn or MgB$_2$.}

\subsection{Critical currents and thermally-activated depinning.}

One may wonder if strong pinning of vortices can reduce their thermal fluctuations and increase $J_c$ and $H^*$. For the cuprates, $J_c$ has been pushed almost to the limit by addition of oxide nanoparticles (for example, Y$_2$O$_3$ or BaZrO$_3$) into YBa$_2$Cu$_3$O$_{7-x}$ films. Such pinning nanostructures can be tuned by varying the shape, size and distribution of oblate or prolate nanoprecipitates, self-assembled chains of nanoparticles or nanorods, typically spaced by 4-10 nm and being 2-4 nm in diameter \cite{y1,y2,y3,y4,y5,y6} to provide the strongest core pining of vortices \cite{ce,blatter,ehb}.

The artificial pinning centers do enhance $J_c$ at low field $H<1$ tesla where $J_c(77 K,0 T) \simeq 5-8$ MA cm$^{-2}$ in thin YBa$_2$Cu$_3$O$_{7-x}$ films at self field can approach $10\%-20\%$ of the GL depairing current density, $J_d=c\phi_0/12\sqrt{3}\pi^2\lambda^2\xi$. Such "designer" pinning nanostructures also increase $J_c$ at intermediate fields most relevant to magnets. For instance, the high values $J_c(0,77K) = 2.7$ MA cm$^{-2}$ and $J_c(5 T,77 K) = 0.1$ MA cm$^{-2}$ were observed on YBa$_2$Cu$_3$O$_{7-x}$ films with Y$_2$O$_3$ nanoparticles \cite{y3}. Meanwhile, recent technological advances have resulted in high $J_c$ in FBS as well, particularly $J_c > 4$ MA cm$^{-2}$ at $4.2$ K  in FBS films \cite{fbs5},  $J_c>0.1$ MA cm$^{-2}$ in Ba(Fe$_{1-x}$Co$-x$)$_2$As$_2$ films and multilayers with BaFeO$_2$ oxide nanorods \cite{ast1,ast2} and $J_c(4.2K)=5 $ MA  cm$^{-2}$ $\simeq 20 \% J_d$ in Ba$_{0.6}$K$_{0.4}$Fe$_2$As$_2$ single crystals irradiated with heavy ions \cite{ast11}.  The FBS conductors and polycrystals have $J_c>0.1$ MA cm$^{-2}$ at 4.2 K and 15-35 tesla \cite{fbswire,ast3,ast4,ast5,ast6,ast7,ast8,ast9,ast10,ast11}, which could make them competitive in power applications.

The high values of $J_c\simeq (0.1-0.3)J_d$ in cuprates and FBS imply that precipitates subdivide vortices into segments of length $\ell \ll\lambda$, nearly as short as the vortex core diameter $2^{3/2}\xi$. As a result, the vortex core and circulating vortex currents get deformed so strongly that the conventional theories of collective pinning \cite{blatter} based on the notion of elementary pinning forces acting on flexible vortex lines with small rigid cores is inadequate. Calculation of $J_c$ then becomes a complex problem which requires self-consistent numerical simulations of the nonlinear GL equations for interacting vortices in the strong pinning potential depending on the shape and spatial distribution of nanoprecipitates \cite{ce,blatter,vp1,vp2,gc}.

To illustrate how far could $J_c$ be increased, I estimate here the upper limit of $J_c$ for strong core pinning by insulating nanoprecipitates of radius $r_0\simeq \xi$ which chop vortex lines into segments of length $\ell\ll\lambda$ as shown in \textbf{Figure 4a}. If the ends of each vortex segments are fixed by strong pins, the complex details of interaction of the vortex core with the defect \cite{ce,blatter,vp1,vp2,gc} become inessential because $J_c$ is determined by the pin breaking mechanism due to cutting and reconnection of vortices \cite{pin1,pin2}. In this case vortices bow out under the action of the Lorentz force ${\bf f_L}=[{\bf J}\times {\bf n}]\phi_0/c$ and escape as the tips of two antiparallel vortex segments at the pin reconnect at the critical current density \cite{brandt}:
\begin{equation}
J_c=\frac{c\phi_0}{8\pi^2\gamma\lambda^2\ell}\ln\frac{\gamma\ell}{2\xi}\left(1-\frac{4\pi r_0^3}{3\eta_c\ell^3}\right).
\label{Jc}
\end{equation}
Here the factor in the parenthesis accounts for the reduction of the current-carrying cross section by randomly-distributed dielectric pins \cite{agpin}, and $\eta_c$ is the percolation threshold which varies from $1/2$ in 2D ($\gamma\gg 1$) to $\simeq 2/3$ in the 3D isotropic limit ($\gamma=1$). Interplay of pinning and current blocking by pins results in the maximum of $J_c(\ell)$ at $\ell_m\simeq (16\pi/3\eta_c)^{1/3}r_0\simeq 3r_0$ in Equation \ref{Jc}. This yields the optimal volume fraction of nanoprecipitates $\eta_m\simeq 9-12\%$ and the maximum $J_c$ of about 20-30$\%$ of $J_d$ \cite{agpin}. Local superconductivity suppression by strains around nanoparticles \cite{strain} can reduce $\eta_m$ in cuprates, but in FBS the optimum $\eta_m\simeq 10\%$ was observed \cite{ast1}. Numerical simulation of vortices pinned by nanoprecipitates was done in Ref. \cite{kosh}.

\begin{figure}
\epsfxsize12pc
\epsfbox{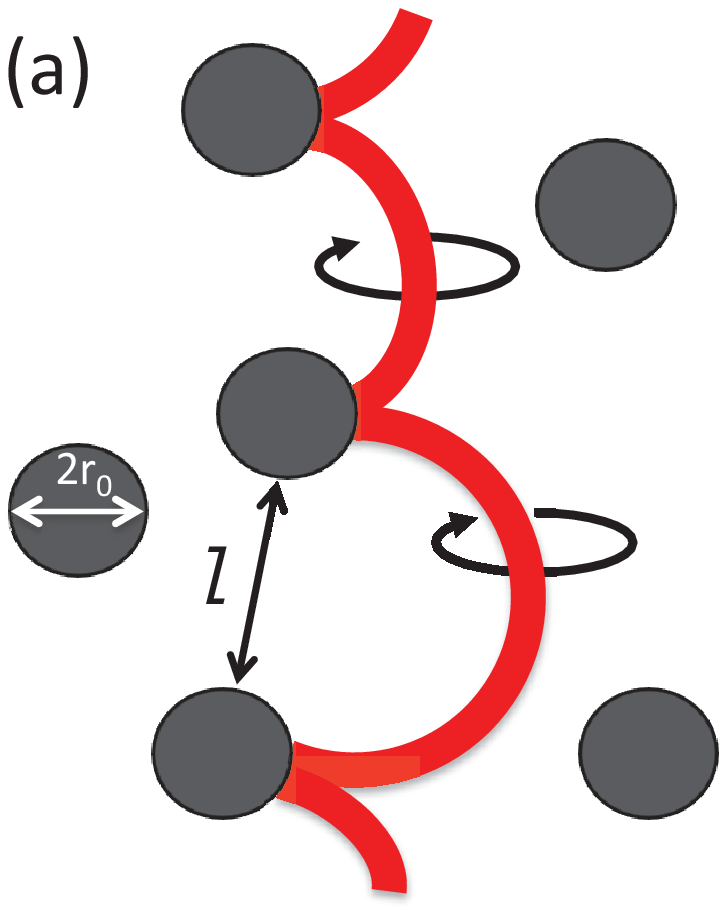}
\hspace
\epsfxsize20pc
\epsfbox{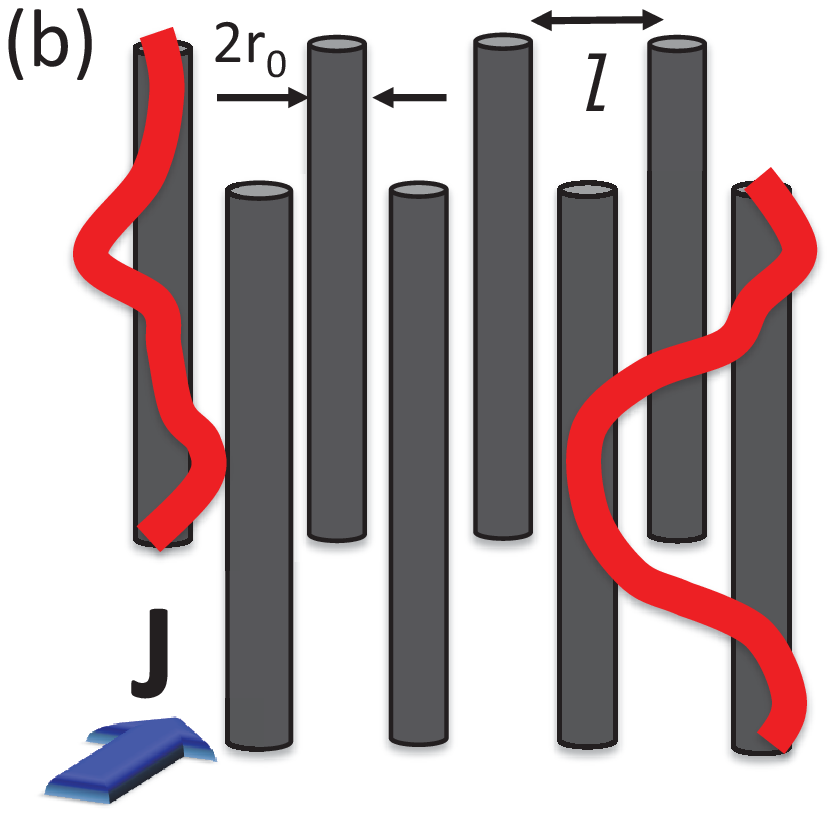}
\caption{Two extreme limits of strong pinning of vortices by dense arrays of randomly-distributed nanoprecipitates (a) or columnar defects parallel to $\mathbf{H}$ (b). Here the red lines show normal vortex cores. }
\label{Figure 4}
\end{figure}

Strong pinning by nanoprecipitates does not assure high $J_c$ at high temperatures and fields where thermally-activated hopping of vortices causes magnetic flux creep \cite{creep} and finite resistivity at $J<J_c$. The significance of giant flux creep for potential RTS applications was pointed out by Tinkham \cite{tinkh} shortly after the discovery of high-$T_c$ cuprates. Flux creep is greatly facilitated by the low vortex line tension which reduces the energy barriers $U_b\sim \varepsilon_0\ell/\gamma$ \cite{brandt} for hopping and reconnection of vortex segments between pins. I illustrate this effect for the most effective pinning by parallel columnar defects of radius $r_0 \sim\xi$ shown in \textbf{Figure 4b}. This model pertains to radiation columnar defects \cite{civale}, or oxide nanorods which have been incorporated into cuprates \cite{y3,y4,y5,y6} and FBS \cite{ast1,ast11} to increase $J_c$. Reduction of $J_c$ due to thermal fluctuations of a vortex trapped by an array of parallel columnar defects was calculated by Nelson and Vinokur \cite{nelvin}. At low temperatures $T<T^*$ and magnetic fields, $J_c\simeq J_d(1-2\eta)$ is close to the depairing limit, but at $T>T^*$ fluctuations significantly reduce $J_c(T)$ due to proliferation of vortex kinks between the columnar defects:
\begin{equation}
J_c\simeq J_d\left(\frac{r_0}{\ell}\right)^3\left(\frac{T^*}{T}\right)^4\left(1-2\pi\frac{r_0^2}{\ell^2}\right), \qquad T^*\simeq\frac{T_c}{1+k_BT_c\gamma/r_0\epsilon_0}.
\label{t*}
\end{equation}
Here the factor $1-2\pi r_0^2/\ell^2$ accounts for the reduction of current-carrying cross section by columnar defects spaced by $\ell$. Interplay of pinning and current blocking yields the maximum $J_m=J_c(\ell_m)$ at the optimum spacing $\ell_m=(10\pi/3)^{1/2}r_0\simeq 3.2r_0$ and the volume fraction of pins $\eta_m\simeq 30\%$ at which $J_m\simeq 0.01J_d(T^*/T)^4$. Here $J_m$ drops to $10^{-4}J_d$ at $T\simeq 3T^*$ and decreases further in a magnetic field.

For YBa$_2$Cu$_3$O$_{7-x}$ with $\lambda_0=150$ nm, $T_c=92$ K, $\gamma=5$ and $r_0=5$ nm, Equation \ref{eps} yields $\varepsilon_0 = 816$ K nm$^{-1}$ and $T^*=83$ K, indicating that columnar pinning remains effective at $77$ K and $J_c\simeq J_d(1-2\eta)$ where $J_d\simeq 300(1-T/T_c)^{3/2}\simeq 20$ MA cm$^{-2}$. The destructive effect of anisotropy is apparent for Bi$_2$Sr$_2$Ca$_2$Cu$_3$O$_x$ with $\lambda_0=200$ nm, $T_c=110$ K and $\gamma=50$ for which $\varepsilon_0=560$ K nm$^{-1}$ and $T^*=37$K. At $77$ K the optimum $J_m\simeq 2(1-T/T_c)^{3/2}(T^*/T)^4\simeq 0.018$ MA cm$^{-2}$ is some four orders of magnitude smaller than for YBa$_2$Cu$_3$O$_{7-x}$.

For RTS with $T_c=400$ K, $\lambda_0=200$ nm, $\lambda/\xi=100$, and a moderate anisotropy $\gamma=7$, Equation \ref{t*} yields $T^*=200$ K and $J_m=2(1-T/T_c)^{3/2}(T^*/T)^4\simeq 0.05$ MA cm$^{-2}$ at 300 K. Moreover, if $k_BT_c\gamma\gg r_0\varepsilon_0$, the crossover temperature $T^*$ approaches the maximum value $T^*_m=r_0\epsilon_0/\gamma k_B$, no matter how high $T_c$ may be. Such RTS would be hardly suitable for magnet applications at 300 K, given that these estimates were made for the strongest pinning by columnar defects at low fields for which the spacing between vortices is much greater than the pin spacing, $H\ll H_m=\phi_0^2/\ell_m^2=3\phi_0/10\pi r_0^2 = 7.6$ tesla. Columnar defects also results in a highly anisotropic $J_c$ with respect to the orientation of ${\bf H}$, although the anisotropy of $J_c$ and the effect of vortex fluctuations can be reduced by combination of columnar and point defects \cite{y3,y4,y5,y6} or by splay distribution of columnar defects \cite{splay}.

\subsection{Grain Boundaries}

 Grain boundaries appear during crystallization of the material and form a 3D network which can play a dual role: it can both pin vortices and block global current.  In Nb$_3$Sn, Nb-Ti, and MgB$_2$ grain boundaries pin vortices and increase $J_c$ \cite{ce,gbfbs} since the Josephson critical current density $J_g$ across GBs is larger than $J_c$. However, in cuprates and FBS, the current-blocking role of GBs dominates because $J_g$ drops exponentially as the misorientation angle $\theta$ between the neighboring crystallites increases:
\begin{equation}
J_g=J_0\exp(-\theta/\theta_0).
\label{jg}
\end{equation}
Here $J_0$ is limited by pinning of vortices in the grains, and $\theta_0\simeq 3-5^\circ$ for the cuprates and $\theta_0\simeq 7-9^\circ$ for FBS \cite{gbhts,gbfbs}.

\begin{figure}
\epsfxsize15pc
\centerline{\epsfbox{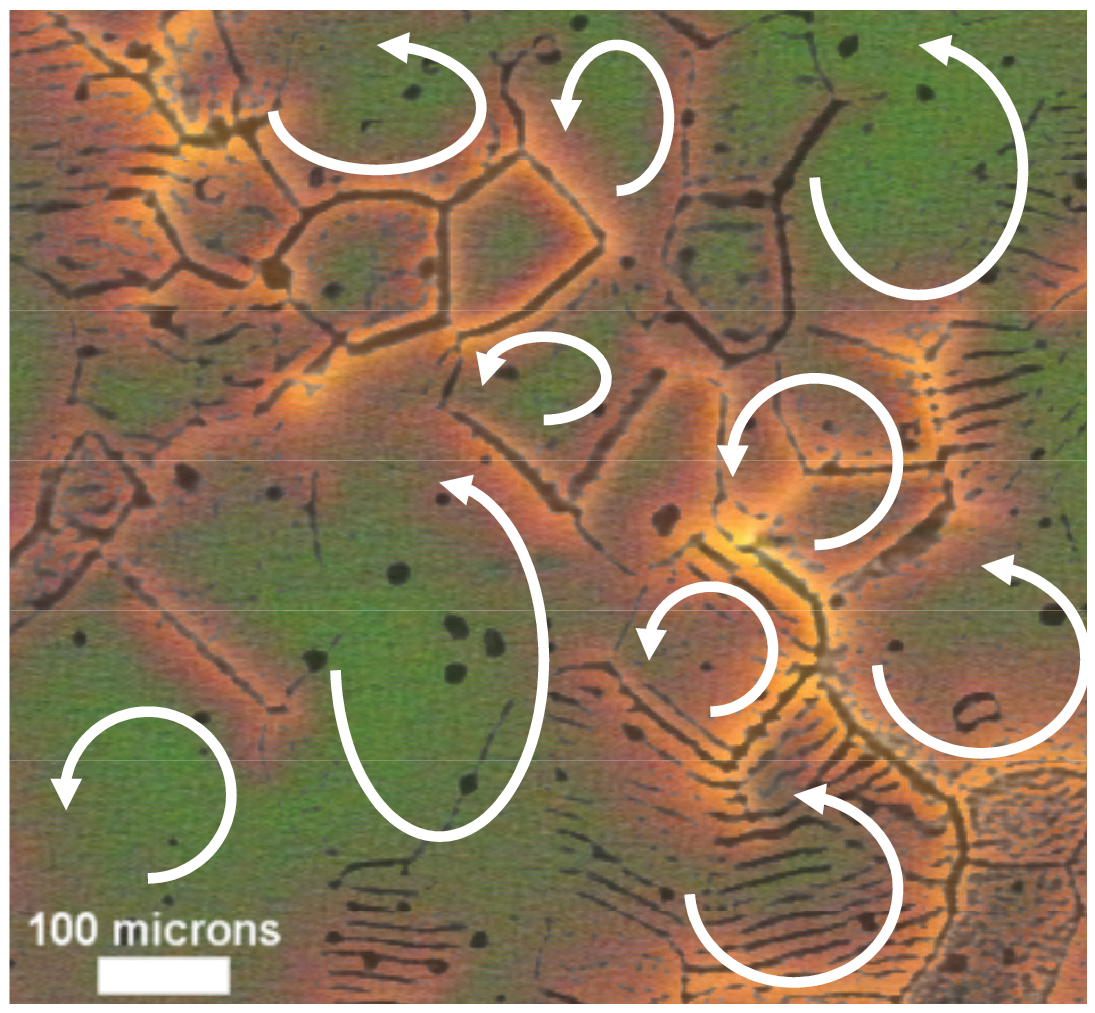}}
\epsfxsize15pc
\centerline{\epsfbox{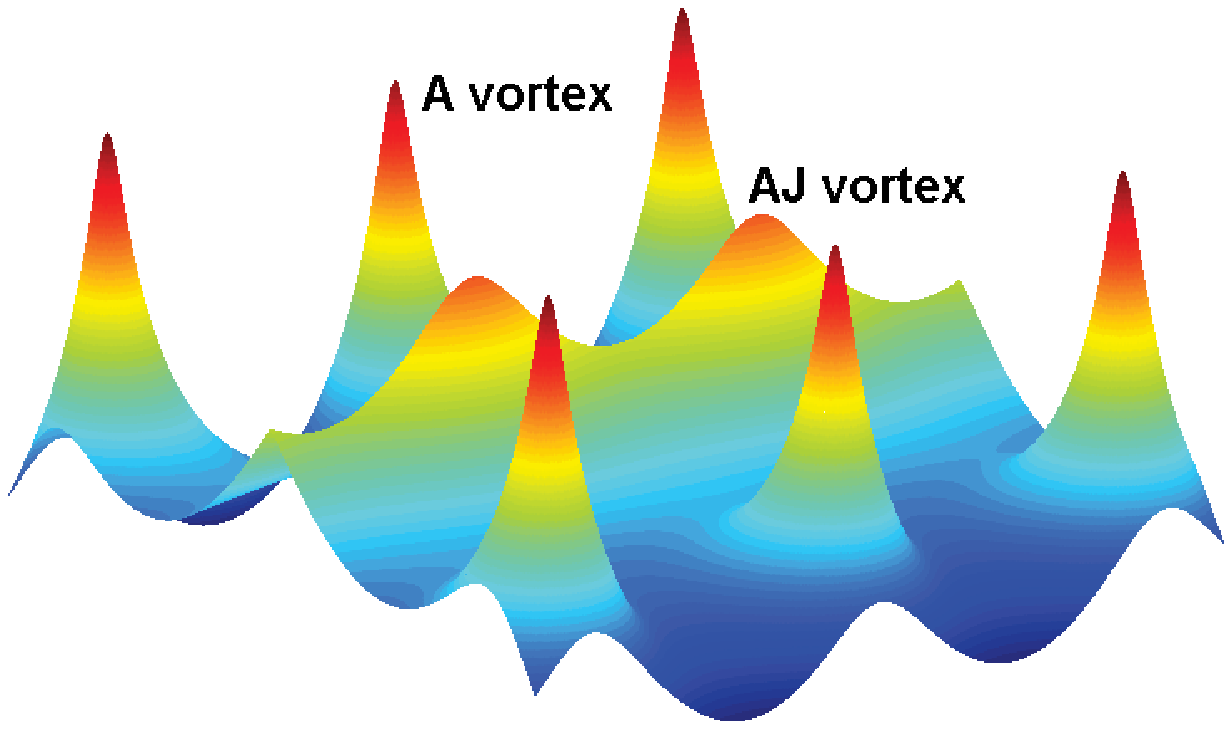}}
\caption{Top: Magnetic granularity in a YBa$_2$Cu$_3$O$_{7-x}$ polycrystal revealed by magneto-optical imaging. Here the yellow contrast shows preferential penetration of magnetic flux along the network of grain boundaries (black lines). As a result, instead of flowing along the conductor, current breaks into weakly connected current loops (white arrows) circulating in the grains. Data by A.A. Polyanskii reproduced from Refs. \cite{natap,agnat}. Bottom: Local field distribution near GB which shows bulk Abrikosov (A) vortices in the grains and mixed Abrikosov-Josephson (AJ) vortices along GB calculated using the approach of Ref. \cite{gc}.}
\label{Figure 5}
\end{figure}

Weak-linked GBs cause magnetic granularity by caging vortices in the grains so that $J_g$ across the GB is smaller than the depinning $J_c$ in the grains. In this case only a fraction of $J_c$ can pass through GBs while most of the current circulates in the grain as shown in \textbf{Figure 5}. The global $J_c$ is limited by pinning of vortices in the grains, and by percolative motion of mixed Abrikosov-Josephson vortices along the GB network \cite{gc}. High $J_g \sim 0.1$ MA cm$^{-2}$ in FBS polycrystals have been achieved even without using biaxial texturing.  For instance, multifilamentary FBS wires with  $J_c\simeq 10-30$ kA cm$^{-2}$ at 4.2 K  and low field \cite{fbswire,ast3,ast4,ast5,ast6,ast7}, and FeSe$_{0.5}$Te$_{0.5}$ coated conductors with $J_c>0.1$ MA cm$^{-2}$ in a wide field range $0<H<30$ tesla at 4.2 K  \cite{ast9} have already been developed.

Grain boundaries in FBS are often coated by non-superconducting amorphous layers \cite{gbfbs,ast8} which aggravate their weak link behavior. Assuming that technological advances will ameliorate porosity, chemical nonstoichiometry and other extrinsic limitations of $J_c$, one can pose a question if ideal clean GBs become intrinsic weak links due to the common features of cuprates and FBS discussed in the Introduction.  Indeed, the GB problem in cuprates cannot be just ascribed to their short coherence lengths, since $\xi_0\simeq 2-3$ nm in Nb$_3$Sn is smaller than $\xi\simeq 4$ nm in YBa$_2$Cu$_3$O$_{7-x}$ at $77$ K. Nor are the d-wave pairing, high $T_c$ or multiband superconductivity sufficient to suppress $J_g$. For instance, MgB$_2$ with $T_c = 40$ K higher than $T_c$ of many FBS does not have the GB problem \cite{gbfbs}.

The proximity of superconductivity to AF states, small $E_F$ and poor electron screening result in greater sensitivity of superconducting properties of FBS and cuprates to small shifts of the chemical potential at GBs. As follows from \textbf{Figure 3}, any local reduction of $x$ may cause superconductivity suppression due to precipitation of the AF phase at the GBs or other crystal defects. Here small $E_F$ and $\xi_0$, and large $l_{TF}$ aggravate the suppression of superconductivity at GBs due to strains and space charge near GB dislocation cores \cite{gbhts,gurpash} or impurity segregation (Cottrell atmospheres) \cite{song}. Because of the large $\partial T_c/\partial\epsilon_{ij}\simeq 200-300$ K even in the optimally-doped cuprates, the strain tensor $\epsilon_{ij}(r)\propto b/2\pi r$ around dislocation cores or can locally suppress superconductivity \cite{gurpash}, where $b$ is the Burgers vector. In FBS $T_c$ also degrades as the bond angle $\alpha ({\bf r})$ in the Fe-As tetrahedrons deviates from $109.5^\circ$ \cite{fbs1,fbs2,fbs3,fbs4,fbs5}. Thus, tilt and shear distortions of $\alpha ({\bf r})$ at the GB dislocation cores may depress superconductivity even at low-angle GBs.

Space charge around GB may be induced by either atomic charges at dislocation cores common in many complex compounds and cuprates, anharmonic expansion of the lattice due to alternating strains around dislocations \cite{gbhts,gurpash} or broken atomic bonds at GB \cite{graser}. The electric potential across the GB, $\varphi_0(x)=(2\pi ql_{TF}/sd\epsilon)\exp(-|x|/l_{TF})$ is proportional to the dislocation charge $q$ per ab plane, where $s$ is the spacing between the ab planes, $d = b/2\sin(\theta/2)$ is the spacing between the GB dislocations,  and $\epsilon$ is the lattice dielectric constant.  The electron band bending at charged GBs \cite{gbhts}, shifts the GB potential by $\varphi_0 = (4\pi ql_{TF}/sb\epsilon)\sin(\theta/2)$.  For $s\simeq b\simeq 0.5$ nm, $l_{TF} = 0.2$ nm, $\varepsilon = 10$, $\theta = 20^o$, and $q$ equal to the electron charge $e$, the energy shift $e\varphi_0\simeq 250$ meV is much smaller than $E_F$ of Nb. However, GBs may become antiferromagnetic or dielectric in cuprates or FBS because $e\varphi_0\sim E_F$.  The order parameter $\Delta_0$ at the GB can be evaluated by solving the GL equation \cite{gurpash}:
\begin{equation}
\Delta_0=\frac{\Delta}{\sqrt{1+\Gamma^2}+\Gamma} , \qquad \Gamma=\frac{2^{3/2}\pi qel_{TF}^2}{sb\xi_0T_c\sqrt{\tau}}\left[\frac{\partial T_c}{\partial\mu}\right]\sin\frac{\theta}{2},
\label{gb}
\end{equation}
where $\Delta$ is the order parameter in the bulk, and $\tau=1-T/T_c$. The pairbreaking parameter $\Gamma$ is amplified by short $\xi_0$, long $l_{TF}$ and large $\partial T_c/\partial\mu\sim T_c/E_F \simeq  0.4-0.7$ K/meV, all of which are characteristic of cuprates and FBS \cite{hts1,fbs1,fbs2,fbs3}.  The suppression of $\Delta_0$ increases at higher $T$, resulting in $J_g\propto (T_c-T)^2$ characteristic of SNS Josephson junctions \cite{book}. The criterion $J_g\ll J_d$ that the GB behaves as a Josephson junction is different from the "practical" definition of $J_g<J_c$. The latter occurs if pinning in the grains is strong enough, so the fact that GBs do not manifest themselves as current-limiting defects in well-connected FBS with $J_c\sim 0.1$ MA cm$^{-2}$ does not mean that GBs are not Josephson junctions, given the scale of $J_d\simeq 120 $ MA cm$^{-2}$ at $T=0$ for $\xi_0=2$ nm and $\lambda_0=200$ nm.

The sign of dislocation core charges is essential: if $q$ has the same sign as the charge of dominant carriers, the underdoped GB facilitates suppression of superconductivity and nucleation of the AF phase \cite{gbfbs}. If dislocation cores and carriers have opposite charges, the GB becomes overdoped (if the grains are optimally doped), which can also depress superconductivity, but without the precipitation of AF phases. In multiband superconductors (like FBS), a charged GB may deplete the carrier density in electron bands and increase it in hole bands, or the other way around. Band bending at GBs is more pronounced in underdoped states as screening becomes less effective. Segregation of impurities in the combined strain and electric fields of charged GB dislocations occurs in a highly inhomogeneous manner as impurities smaller than the host atoms cluster in compressed regions of the lattice and avoid dilated regions at the dislocation cores. Such Cottrell atmosphere of impurities form in the dipole strain fields of GB dislocations, oscillating along GB and decaying across the GB \cite{song}.

The GL models incorporating the mechanisms discussed above \cite{gurpash} capture the strong decrease of $J_g(\theta)$ in cuprates and FBS for low-angle GBs with $\theta < 10-15^\circ$ in which the dislocation cores do not overlap. A microscopic model of $J_g(\theta)$ for high-angle GBs based on self-consistent simulations of the GB structure combined with the Bogolubov-deGennes equations was proposed by Graser et al. \cite{graser}.  This model can explain the exponential decrease of $J_g(\theta)$ with $\theta$ although it does not take into account the Mott-Hubbard physics \cite{hts2} responsible for the phase diagram shown in \textbf{Figure 2}.

Electromagnetic granularity can be ameliorated by overdoping GBs with impurities, for example, by Ca overdoping of YBa$_2$Cu$_3$O$_{7-x}$ at the expense of reduced $T_c$ in the grains \cite{gbhts}. The GB problem in FBS appears less severe than in cuprates, which may enable FBS multifilamentary wires with $J_c\sim 0.1$ MA cm$^{-2}$ for magnet applications at 4.2 K. Indeed, a polycrystalline  (Ba$_{0.6}$K$_{0.4}$)Fe$_2$As$_2$ wire at 4.2 K can carry $J_c\simeq 0.1$ MA cm$^{-2}$ in self field and $\simeq 10$ kA cm$^{-2}$ at 15 tesla \cite{ast10}. Multifilamentary wires of K-doped BaFe$_2$As$_2$ or SrFe$_2$As$_2$ can carry $J_c\simeq 20$ kA cm$^{-2}$ in self field at 4.2K, but $J_c(T,B)$ decreases with field rather rapidly at $T>10$ K \cite{fbswire}.  It remains to be seen if high $H_{c2}$ and $J_c$ at $20-30$ K will suffice to develop FBS conductors in the important field range $H \simeq 20-30 $ tesla where they could have advantage over Nb$_3$Sn.

\section{Room temperature superconductivity.}

In addition to the effect of fluctuations of vortices on current transport in a magnetic field, thermal fluctuations of the order parameter $\Psi({\bf r})$ at $H=0$ reduce the critical temperature $T_c$ at which the macroscopic phase coherence sets in. As a result, $T_c$ becomes lower than the pairing temperature $T_p$ below which the superconducting correlations cause the formation of Cooper pairs \cite{kivels1}. There have been many suggestions to increase $T_p$ by excitonic mechanisms in quasi-1D or sandwich structures, \cite{little,ginzburg, bardeen}, or by magnetic excitations in doped antiferromagnets  \cite{hts2,hts3,hts4,hts5,hts6,hts7,fbst1,fbst2,pickett,geballe,chu}. Utilizing high frequency electron or spin excitations $\hbar\Omega\sim E_F$ of the same Fermi liquid which becomes superconducting is the main challenge for these theories which can no longer take advantage of the small Migdal parameter $\hbar\omega_D/E_F$ of the Eliashberg theory in which the low-frequency pairing glue (phonons) exists independently of the electrons.

The problem of RTS materials requirements could be re-formulated from a different perspective which is less sensitive to pairing scenarios. Suppose that there is a microscopic mechanism which provides a high pairing temperature $T_p$ renormalized by thermal and quantum fluctuations with wavelengths $<\xi$. Let us estimate how $T_c$ gets reduced by fluctuations on larger scales $>\xi$. Generally, the phase coherence is destroyed by thermal fluctuations at temperatures for which $k_BT$ becomes of the order of the condensation energy $H_c^2V/8\pi$ in the correlated volume $V_3=\xi^2\xi_c$ of an anisotropic 3D superconductor or $V_2=s\xi^2$ in the 2D layered limit. Hence, $T_c$ is determined self-consistently by the equations
\begin{eqnarray}
k_BT_c\simeq g_2H_c^2s\xi^2/8\pi, \qquad 2D
\label{2d} \\
k_BT_c\simeq g_3H_c\xi^2\xi_c/8\pi, \qquad 3D
\label{3d}
\end{eqnarray}
where the right-hand sides depend on both $T_p$ and $T_c$, and the numerical factors $g_2$ and $g_3$ will be discussed below. To keep things simple, I use the GL result $H_c^2=\phi_0^2/8\pi^2\lambda^2\xi^2$ in which $\lambda^2 = \lambda_0^2(1 - T^2/T_p^2)^{-1}$ and $\xi^2 = \xi_0^2(1 - T^2/T_p^2)^{-1}$  diverge at $T_p$. Then Equations \ref{2d} and \ref{3d} define $T_c$ in terms of $T_p$ and the maximum temperature $T_m$ at which the phase coherence can exist:
\begin{eqnarray}
T_c=-\frac{T_p^2}{2T_m}+\left(\frac{T_p^4}{4T_m^2}+T_p^2\right)^{1/2}, \qquad T_m=\frac{g_2\phi_0^2s}{64\pi^3k_B\lambda_0^2},    \qquad 2D
\label{t2} \\
T_c=\frac{T_m}{(1+T_m^2/T_p^2)^{1/2}}, \qquad T_m=\frac{g_3\phi_0^2\xi_0}{64\pi^3k_B\gamma\lambda_0^2},    \qquad 3D
\label{t3}
\end{eqnarray}
where $s$ is the total thickness of coupled 2D superconducting layers. For $Gi\ll 1$,  Equations \ref{t2} and \ref{t3} define the narrow Ginzburg-Levanyuk temperature region $T_p-T_c\sim Gi T_p$ of critical fluctuations near $T_c$ \cite{fluct}.

Equations \ref{t2} and \ref{t3} show that $T_c$ can be much smaller than $T_p$ if $Gi>1$. In 2D the critical temperature $T_c$ monotonically increases with $T_p$, approaching the limiting value $T_m$ independent of $T_p$. However, using the BCS result $\xi_0=\hbar v_F/2\pi k_BT_p$ in Equation \ref{t3} yields a maximum $T_c$ at $T_p \sim E_F/k_B$ and $\xi_0\sim r_s$. Since $\xi_0<r_s$ would imply no phase coherence due to overlapping Cooper pairs, I only discuss here the case of $k_BT_p <E_F$. The relation between $T_c$ and $T_p$ resembles the relation between $H^*$ and $H_{c2}$ discussed above:  $H^*\rightarrow H_{c2}$ for weak vortex fluctuations and $H^*(T)$ independent of $H_{c2}$ for strong fluctuations. The limiting value $T_m\propto n_s\propto\sum_i(n_i/m_i)$ is determined by the partial carrier densities $n_i$ and the electron/hole masses $m_i$ in all bands relevant to superconductivity.

The importance of phase fluctuations for reducing $T_c$ in superconductors with low $n_s$ was recognized shortly after the discovery of high-$T_c$ cuprates \cite{kivels1}. Much attention has been given to the renormalization of $n_s$ by nonlinear phase fluctuations due to thermally-activated proliferation of entangled vortex loops \cite{vortf1,vortf2}, or to the duality of quantum fluctuations in strongly correlated systems and weakly coupled quantum gravity in $D+1$ dimensions \cite{sachdev}. The 2D manifestation of vortex fluctuations is the Berezinskii-Kosterlitz-Thouless (BKT) unbinding of vortex-antivortex pairs in thin films or layered materials with weak Josephson interaction between superconducting planes \cite{bkt}.

\begin{figure}
\epsfxsize15pc
\centerline{\epsfbox{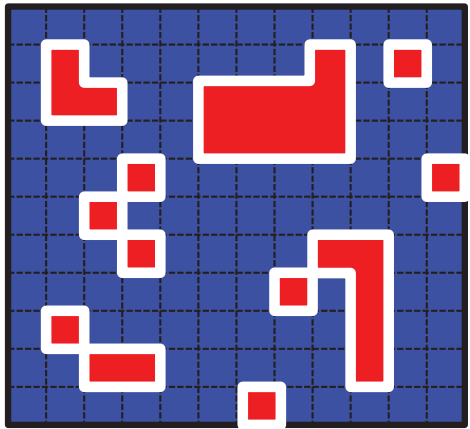}}
\caption{A snapshot of fluctuating 2D "phase polycrystal". The dashed lines show $2^{3/2}\xi\times 2^{3/2}\xi$ cells of equilibrium order parameter $\Psi$ with either positive (blue) or negative (red) signs. The antiphase cells with different signs of $\Psi$ are separated by domain walls (white) with the energy $2J$ per one side of the cell.  This model can be mapped onto the ferromagnetic Ising model with the Hamiltonian $\hat{H}=-J\sum_{\mathbf{mn}}\sigma_{\mathbf{m}}\sigma_{\mathbf{n}}$ where the summation goes over nearest neighbors. Here the "spin" variable $\sigma_\mathbf{n}=\pm 1$ describes the sign of $\Psi$ in the cell with the coordinate $\mathbf{n}$. If the free energy is counted from the fully ordered state (blue), then the energy of the domain wall equals $2J$.  }
\label{Figure 6}
\end{figure}

Here I only estimate an upper bound of $T_c$ using a transparent model in which only one kind of fluctuations, namely antiphase domains with opposite signs of $\Psi$ are taken into account, but inhomogeneous phase fluctuations causing pairbreaking currents are disregarded. In this model a superconductor is subdivided into minimal $2^{3/2}\xi\times 2^{3/2}\xi\times 2^{3/2}\xi_c$ cells in which $\Psi({\bf r})$ can only take the equilibrium value of $\Psi_0$ with either positive or negative signs. Neighboring cells with opposite signs of $\Psi$ are separated by the domain walls, $\Psi(x)=\Psi_0\tanh(x/\xi\sqrt{2})$ of width $\simeq \xi$ and the energy $2J=A\sigma$, where the area $A=2^{3/2}s\xi$ for 2D and $A=8\xi^2$ for 3D, and $\sigma = H_c^2\xi\sqrt{2}/3\pi$ is the surface energy of the domain wall calculated from the GL theory \cite{pslip}.

The thermodynamics of such fluctuating "phase polycrystal" with all possible configurations of domain walls depicted in \textbf{Figure 6} can be mapped onto the ferromagnetic Ising model with the exchange interaction energy $J$, and up and down spins corresponding to $+$ and $-$ signs of $\Psi(x,y)$ in each cell. This analogy suggests a transition temperature $T_c\sim J(T_c)/k_B$ above which the number of positive and negative domains is the same, so the superconducting phase coherence is lost. At $T<T_c$ an ordered phase-coherent state appears. This model overestimates $T_c$ since it disregards fluctuations due to pairbreaking vortex loops.

Substituting $J=2s\xi^2H_c^2/3\pi$ into the Onsager solution $k_BT_c=2J/\ln(1+\sqrt{2})=2.67J(T_c)$ yields Equation \ref{t2} with $g_2=g_{DW}\approx 12$. Notice that Equation \ref{t2} with $g_{BKT}=2\pi\approx 0.52g_{DW}$ reduces to the condition of the BKT transition, $k_BT_c=\phi_0^2s/32\pi^2\lambda^2(T_c)$, thus vortex and domain wall fluctuations give contributions of the same order of magnitude. In real layered superconductors, long-range Josephson strings between fluctuating vortex pancakes on different superconducting planes increase $g_2$, making vortex fluctuations more energy costly \cite{blatter}.

For isotropic superconductors, using $J=4\xi^2\sigma$ in the equation $k_BT_c\approx 4.5J(T_c)$ of the 3D Ising model \cite{ising} yields Equation \ref{t3} with $g_3\approx 68$. Reduction of $T_c$ by vortex fluctuations would be of the same order of magnitude as for the phase polycrystal. Therefore, Equations \ref{t2} and \ref{t3} with $g_2\approx 12$ and $g_3\approx 68$ in which vortex fluctuations are disregarded could be used to estimate the upper bound of $T_c$.  Even this rough estimate can put essential constraints on the normal state electronic parameters of RTS operating at $300$ K.

The behavior of $T_m\propto\lambda_0^{-2}\propto n_s$ in Equation \ref{t2} is consistent with the Uemura plot \cite{hts1,hts2,hts3,hts4}. For typical values of $\lambda_0 = 200$ nm and the thickness $s=0.3$ nm of a single Cu-O plane, Equation \ref{t2} yields $T_m = 132$ K. For $T_p = 400$ K, Equation \ref{t2} gives $T_c = 120$ K, while a naive BKT estimate for a single layer yields $T_c\simeq 67$ K, indicating that RTS would be unlikely among extremely layered compounds with electronic parameters of cuprates and FBS. Furthermore, if $s$ in Equation \ref{t2} is the thickness of several coupled Cu-O planes, $T_m$ increases with the number of planes until $s$ reaches $\simeq \xi_c$ in the stack of these planes, and $T_m$ levels off. This behavior is qualitatively consistent with the dependence of $T_c$ on the number of layers observed in HgBa$_2$Ca$_{n-1}$Cu$_n$O$_{3n+2}$ compounds \cite{hts1}. This model in which superconducting properties of Cu-O planes do not change with $n$ cannot explain the observed maximum in $T_c$ at $n=3$, yet it suggests that $T_m$ would roughly triple at $n=3$, giving $T_m=396$ K and $T_c=246$ K at $T_p=400$ K.

The restrictions imposed by fluctuations become weaker for less anisotropic materials with higher superfluid density and $\lambda_0\simeq 100$ nm characteristic of such conventional superconductors as Nb$_3$Sn and MgB$_2$. It is the case when just a two-fold decrease of $\lambda_0$ could make a big difference for RTS applications. Equally important is the effect of mass anisotropy in Equation \ref{t3}. For instance, an RTS analog of YBa$_2$Cu$_3$O$_{7-x}$ with $T_p=400$ K, $\lambda_0=150$ nm, $\lambda_0/\xi_0=100$, and $\gamma=5$ would have $T_m=1324$ K and the upper bound of $T_c=383$ K.

\section{Conclusion and Outlook}

The development of high-$T_c$ cuprates and FBS has resulted in the evolution of parameters of merits for applications at high temperatures and magnetic fields, from the pairing-limited $T_p$ and $H_{c2}$ to the parameters mostly controlled by thermal fluctuations and transparency of grain boundaries. This shift of perception reflects the recognition of the importance of thermal fluctuation of vortices at high $T$ and $H$ in layered strongly-correlated superconductors. Those unconventional superconductors with low Fermi energies, non-phonon pairing, and competing AF orders, can have $T_c$ up to 132 K for cuprates and up to 55 K for FBS, and very high $H_{c2}(0)>100$ tesla. However, making these materials useful for power or magnet applications at 30-77 K requires a lot of innovative materials tuning and expensive technological compromises.

For instance, the coated conductor shown in {\bf Figure 2} utilizes only a few $\%$ of the current-carrying cross-section, so $J_c$ of the superconducting film must be pushed to the limit by incorporating dense arrays of nanoprecipitates spaced by $\ell \sim 10$ nm. This does increase $J_c$ at low $T$ and $H$, but not necessarily $H^*$ at 77K or higher temperatures. Indeed, because of low vortex line tension $\varepsilon_l(T) \sim 1-10$ K nm$^{-1}$, vortex segments of length $\sim k_BT/\epsilon_l\sim 10-100$ nm $>\ell$ can hop and reconnect at neighboring pins. In turn, low energy barriers $U_b\sim\ell\epsilon_0\gamma^{-1}\sim k_BT$ for cutting and reconnection of soft vortex segments \cite{ehb} between nanoprecipitates cause thermally-activated flux creep and electrical resistance at $J<J_c$, no matter how strong pinning may be. Thus, the bad combination of low $\varepsilon_l(T)$ and  weak-linked GBs would make it hard to develop competitive coated conductors for high-field applications at $T>77$ K using superconductors in which $\varepsilon_l(T)$ is lower than $\varepsilon_l$ for YBa$_2$Cu$_3$O$_{7-x}$.

As far as the search for new materials is concerned, mechanisms which provide high pairing temperature may not necessarily result in RTS which can actually be used at 300 K. In fact, it may be beneficial to explore new materials means of reducing fluctuations in cuprates or FBS, for example, by reducing the electronic anisotropy with chemical substitutions or using contact with normal metals \cite{kivel} or managing optimal inhomogeneities at the nanoscale \cite{inhom}. Certainly, searching for new materials with higher superfluid density and lower anisotropy than in cuprates or FBS, even at the expense of lower $T_c$, could be very fruitful, as the success of MgB$_2$ has shown. Curiously, a viable RTS may have to satisfy one of the Matthias criteria (cubic symmetry is best), but not the one which requires peaks in the density of states which would increase $m$ and $\lambda_0$.  The following temperature regions for applications of different materials can be identified:
\begin{itemize}
\item{\textbf{Helium temperatures.} All existing superconductors with high $H_{c2}$ can be used. Conventional parameters of merits are applicable, fluctuations are negligible, and the cheapest and most technological materials win.}
\item{\textbf{Intermediate temperatures} (20-40 K). High $H_{c2}(T)$ and $H^*(T)$ of MgB$_2$, FBS and Bi-based cuprates at 20-30 K can make them useful in magnet applications. Effect of fluctuations is noticeable in FBS and Bi-based cuprates, but the GB problem in FBS appears not as bad as in cuprates.}
\item{\textbf{Nitrogen temperatures}. Because of strong vortex fluctuations, only the least anisotropic cuprate YBa$_2$Cu$_3$O$_{7-x}$ can be used in magnets. Resolving the GB problem requires coated conductor technologies.}
\item{\textbf{Room temperatures}. Performance of RTS will likely be limited by fluctuations. The higher the operating temperature, the less relevant $T_p$ and $H_{c2}$ become. Even the  optimistic conditions of RTS applications impose such restrictions on the carrier density and electronic anisotropy, that they may only be satisfied for the normal state materials parameters
similar to those of conventional, weakly anisotropic superconductors.}
\end{itemize}

So what can all this do to the nice idea of RTS levitating trains, power transmission lines or high field magnets? To this end, the answer may be irrelevant because the use of any superconductor requires a cooling system, not only to keep the temperature below $T_c$ but also to provide the necessary cryogenic stability against thermal quench or to manage losses in alternating fields \cite{wilson}.  The use of liquid nitrogen refrigeration is a cheap and environmentally safe solution for any material with $T_c>77$ K which would also resolve most of the fundamental problems of genuine RTS applications. For instance, a superconductor with $T_c=240$ K could be a revolutionary material for power and magnet applications at 77 K if it does not have problems with low $H^*$, current-blocking GBs, and the environmental safety of chemical components.

There is no fundamental reason why RTS cannot exist, but as the operating temperature increases, the materials parameter space where applications are possible shrinks rapidly, so the big question is to what extent properties of RTS can be tuned to make them useful at high $T$ and $H$. Addressing this question will require broad investigations of the physics and materials science of unconventional superconductivity, particularly the effect of impurities, which is far more complex than for BCS superconductors \cite{imp1,imp2}. For instance, it remains unclear if nonmagnetic and magnetic impurities can be used to reduce the electronic anisotropy of superconductors and weaken the effect of fluctuations. It is an area of research in which much fundamental physics and materials science gets intertwined, but clarifying these important issues would be instrumental to make unconventional superconductors useful if the next breakthrough in the discovery of higher $T_c$ materials will indeed occur.

\end{document}